\documentclass[12pt,onecolumn]{IEEEtran}
\usepackage{amsfonts}
\IEEEoverridecommandlockouts

\ifCLASSINFOpdf
\else
\fi
\usepackage{epstopdf}
\usepackage{multirow}
\usepackage{cite}
\usepackage{stfloats}
\usepackage{color}
\usepackage{epsfig}
\usepackage{graphicx}
\usepackage{psfig}
\usepackage{epsf}
\usepackage{slashbox}
\usepackage{float}
\usepackage{amssymb }
\usepackage[cmex10]{amsmath}
\begin{document}

\title{Optimal Combining and Performance Analysis for  Two-Way EH Relay Systems with TDBC Protocol }
\author{Liqin Shi, Yinghui Ye, Xiaoli Chu,~\IEEEmembership{Senior Member,~IEEE,} Yangyang Zhang, and Hailin Zhang,~\IEEEmembership{Member,~IEEE}%\vspace*{-3.1em}
\thanks{Liqin Shi, Yinghui Ye, Yangyang Zhang and Hailin Zhang  are with the State Key Laboratory of Integrated Service Networks, Xidian University, Xi'an, China (e-mail:{ liqinshi@hotmail.com}, { connectyyh@126.com}, {zyy\_xidian@126.com}, { hlzhang@xidian.edu.cn}).
Xiaoli Chu (e-mail:
x.chu@sheffield.ac.uk) is with the Department of Electronic and Electrical Engineering,
University of Sheffield, U.K.}
\thanks{This work was supported by the scholarship from China Scholarship Council, the National Natural Science Funding of China under grant 61701364,  the Natural Science Funding of Shannxi Province under grant 2018JM6019 and the 111 Project of China under grant B08038.}
}
\markboth{IEEE Wireless Communications Letter, No. XX, MONTH YY, YEAR 2018}
{Shi\MakeLowercase{\textit{et al.}}: Outage Performance Optimization for SWIPT Enabled Three-Step Two-Way DF Relaying}
\maketitle

\begin{abstract}
In this paper, we investigate a simultaneous wireless information and power transfer (SWIPT) based two-way decode-and-forward (DF) relay
network, where time switching (TS) is employed for SWIPT and time division broadcast (TDBC) is employed for two-way relaying. We focus on the design of a combining scheme that decides how the relay combines the signals received from two terminals through a power allocation ratio at the relay. We formulate an optimization problem  to minimize the system outage probability and obtain the optimal power allocation ratio in closed form.
%We propose an optimal combining scheme by optimizing the power allocation ratio at the relay, and obtain the closed-form expression for its optimal solution.
For the proposed optimal combining scheme, we derive the expression for the system outage probability.
Simulation results verify our derived expressions and show that the proposed scheme achieves a lower system outage probability than the existing schemes.
\end{abstract}
\IEEEpeerreviewmaketitle
%\vspace*{-10pt}
\section{Introduction}
\IEEEPARstart{O}{wing} to its high spectrum efficiency, two-way relaying has been deemed an integral part of Internet of Things \cite{7744827}. It allows two terminals to exchange their information through an intermediate relay using either the multiple access broadcast (MABC) or time division broadcast (TDBC) protocol. However, in an energy-constrained wireless sensor network, the intermediate relay  is likely to  have a limited battery capacity and would be unwilling to assist the terminals \cite{7744827}. To solve this problem,  simultaneous  wireless information and power transfer (SWIPT) based two-way relaying, where the intermediate relay splits or switches the received radio frequency (RF) signal in the power or time domain through power-splitting (PS) or time-switching (TS), has been proposed.
SWIPT based two-way relaying with MABC has been widely studied (see \cite{7831382, 7807356,7876801} and the references therein). {\color{black}For example, the authors in \cite{7831382} investigated the system outage probability of an analog network coding based two-way amplify-and-forward (AF) relay system with multiple-antenna source terminals and a single-antenna energy harvesting (EH) relay. Authors in \cite{7876801} proposed an energy efficient precoding design for SWIPT enabled MIMO two-way AF relay networks.}

{\color{black}Since the TDBC protocol can utilize the direct link between the terminals even when they operate in a half-duplex mode and the operational complexity of TDBC is lower than that of MABC, the study of TDBC in SWIPT based two-way relaying has recently received a lot of attention \cite{7037438,8287997,8361446,2017CL,shi,8377371,8103768,8364583}.}
%ever-increasing attentions  simpler circuit design is required for SWIPT based two-way relaying with TDBC , this paper focuses on the study of .
In \cite{7037438}, the authors studied the achievable throughput of SWIPT based additive AF relaying with TBDC under three wireless power transfer policies.  For SWIPT based multiplicative AF relaying with TBDC, optimal symmetric PS \cite{8287997} and asymmetric PS \cite{8361446} schemes were proposed to minimize the system outage probability.
The authors in \cite{2017CL} and \cite{shi} investigated the outage probability of PS SWIPT based two-way  decode-and-forward (DF) relaying with TDBC under both linear and non-linear energy harvesting models. {\color{black}The authors in \cite{8377371} studied the outage performance of a decode-amplify-forward protocol in  TS SWIPT based two-way relaying.} {\color{black}The secrecy performance, e.g., the intercept probability, of PS SWIPT based two-way  DF relaying with TDBC was investigated in \cite{8103768}. The combination of PS SWIPT based two-way  DF relaying with TDBC and cognitive radio was studied in \cite{8364583} with a focus on the outage probabilities of the primary user and the secondary user.} It has been shown that for TDBC, the combining scheme that combines the two received signals at the relay can significantly  improve the outage performance \cite{5738653}. However, the optimization of combining scheme has not been studied for SWIPT based two-way relaying with TDBC.

In this paper, we propose an optimal combining scheme for a TS SWIPT based two-way DF\footnote{{\color{black}Although AF relaying is simpler than DF relaying, the major drawback of AF relaying is noise amplification at the relay, which may degrade the received signal-to-noise ratio (SNR) at the destination node. DF relaying is a commonly used relaying protocol for eliminating the noise amplification effect.}} relay network, where TDBC is employed.
%a TS scheme is employed to incentivize the relay to cooperate with terminal nodes.
%To be general, we assume that the direct link between two terminal nodes exists.
{\color{black}Our contributions are summarized as follows.
Firstly,
we propose an optimal combining scheme, where the power allocation ratio at the relay is adjustable. Specifically, we formulate an optimization problem to minimize the system outage probability and derive the optimal power allocation ratio in closed form.
Secondly, with the optimal combining scheme, we derive the expression of the system outage probability considering the EH circuit sensitivity.
%Note that the derived analytical expression enables the analysis of the system outage probability as functions of the relay location, the time allocation for energy harvesting, and path loss exponent.
Simulation results are presented to verify the accuracy of the derived expressions and demonstrate the superiority of the proposed scheme in terms of system outage performance.}

%\begin{figure}[!t]
%  \centering
%  \includegraphics[width=0.4\textwidth]{systemmodel6}\\
%  \caption{System model of the SWIPT based DF TWRN}
%\end{figure}
%
%\begin{figure}[!t]
%  \centering
%    \includegraphics[width=0.4\textwidth]{fig0.pdf}
%  \caption{Transmission time-block structure for the proposed dynamic asymmetric scheme}\label{fig.5}
%\end{figure}
%\vspace*{-8pt}
\section{System model}
We consider a TDBC based two-way DF relay network in the presence of a direct link between the terminal node A and the terminal node B, where the \lq\lq harvest-then-forward\rq\rq$ $ strategy is adopted to incentivize the energy-constrained relay R to help with the information transmission between node A and B.
All participating nodes are equipped with a single antenna\footnote{{\color{black} Note that energy harvesting is particularly applicable to wireless sensor networks \cite{7744827}, where it may be difficult for low-cost, small wireless sensor nodes to have multiple antennas. Meanwhile, energy harvesting with a single antenna at each node has also been assumed in many related recent works, e.g., \cite{7037438,8287997,8361446,2017CL,shi,8377371,8103768,8364583}.}}and operate in the half-duplex mode.
All channels are assumed to be reciprocal and undergo independent and identically distributed (i.i.d.) Rayleigh fading \cite{7831382,7807356,7037438,2017CL}.
Let $h_{i}(i\in\{\rm{A},\rm{B}\})$ denote the fading coefficient between $i$ and $\rm{R}$, where $|h_{i}|^2\sim \exp(\frac{1}{\lambda_i})$.
The path loss of the link between $i$ and $\rm{R}$ is given by $d_{i}^{ - \alpha_s }$, where $d_i$ and $ \alpha_s$ are the distance and the path loss exponent of the $i-\rm{R}$ link, respectively. For the direct link between \rm{A} and \rm{B}, the channel model is denoted by $|g|^2d_{t}^{ - \alpha_t }$, where $g$ is the fading coefficient, $d_t$ is the distance between $\rm{A}$ and $\rm{B}$, and $\alpha_t$ is the corresponding path loss exponent.

Relay $\rm{R}$ adopts TS SWIPT, where each transmission block $T$ is divided into four time slots.
{\color{black}During the first time slot of duration $\beta T, \beta\in(0,1)$, the relay harvests energy from the RF signals transmitted by both \rm{A} and \rm{B}. The total harvested energy is given by
\begin{align}\label{1}
{E_{{\rm{total}}}} = \left\{ {\begin{array}{*{20}{c}}
{0,\;\;\;\;\;\;\;\;\;\;{\rm{   }}{P_{\rm{in}}} < {P_{\rm{th}}}}\\
{\beta T\eta {P_{\rm{in}}},{\rm{  }}{P_{\rm{in}}} \ge {P_{\rm{th}}}}
\end{array}} \right.
\end{align}
where ${P_{\rm{in}}}= P\left( |{h_{\rm{A}}{|^2}d_{\rm{A}}^{ - {\alpha _s}} + |{h_{\rm{B}}}{|^2}d_{\rm{B}}^{ - {\alpha _s}}} \right)$ is the received RF power at the relay;} $P$ is the transmit power used by \rm{A} and \rm{B}; ${P_{\rm{th}}}$ is the circuit sensitivity of the energy harvester; and $\eta\in(0,1]$ is the energy conversion efficiency of the energy harvester.
%\begin{figure}
%  \centering
%  \includegraphics[width=0.36\textwidth]{systemmodel.pdf}\\
%  \caption{An illustration of system model.}\label{fig0}
%  \vspace*{-13pt}
%\end{figure}

{\color{black}In the second (third) time slot of duration $\frac{1-\beta}{3} T$,
$\rm{A}$ ($\rm{B}$) transmits its signal $s_{\rm{A}}$ ($s_{\rm{B}}$).} {\color{black}The received signals from $i\;(i\in\{\rm{A},\rm{B}\})$ at relay $\rm{R}$ and terminal $\overline{i}$ are given respectively by
\begin{align}\label{2}
&{y_{ i\rm{R}}} = {h_{ i}}\sqrt {{P}d_{i}^{-\alpha_s}} {s_{ i}} + {n_{ \rm{R}}},
&{y_{ i\overline{i}}} = {g}\sqrt {{P}d_{t}^{-\alpha_t}} {s_{ i}} + {n_{ \overline{i}}},
\end{align}
where $\bar i = \left\{ {\begin{array}{*{20}{c}}
{{\rm{A,   if }}\;i = {\rm{B}}{\rm{ }}}\\
{{\rm{B}},{\rm{  if }}\;i = {\rm{A}}}
\end{array}} \right.$;
$\mathbb{E}\left\{ {{{\left| s_{i} \right|}^2}} \right\} = 1$; ${n_{\rm{R}}} \sim {\rm{{\cal C}{\cal N}}}\left( {0,\sigma _{\rm{R}}^2} \right)$ and ${n_{\overline{i}}} \sim {\rm{{\cal C}{\cal N}}}\left( {0,\sigma _{\overline{i}}^2} \right)$ are the additive white Gaussian noise (AWGN) at \rm{R} and $\overline{i}$, respectively.
Thus, the received SNR at \rm{R} and $\overline{i}$ are given by
\begin{align}\label{3}
%&{\gamma_{ i\rm{R}}} = \frac{{P{{\left| {{h_i}} \right|}^2}d_i^{ - {\alpha _s}}}}{{\sigma _{\rm{R}}^2}},
&{\gamma _{i{\rm{R}}}} = {{P{{\left| {{h_i}} \right|}^2}d_i^{ - {\alpha _s}}} \mathord{\left/
 {\vphantom {{P{{\left| {{h_i}} \right|}^2}d_i^{ - {\alpha _s}}} {\sigma _{\rm{R}}^2}}} \right.
 \kern-\nulldelimiterspace} {\sigma _{\rm{R}}^2}},
&{\gamma_{ i\overline{i}}} ={{P{{\left| g \right|}^2}d_t^{ - {\alpha _t}}} \mathord{\left/
 {\vphantom {{P{{\left| g \right|}^2}d_t^{ - {\alpha _t}}} {\sigma _{\bar i}^2}}} \right.
 \kern-\nulldelimiterspace} {\sigma _{\bar i}^2}}.
 \vspace*{-30pt}
\end{align}}

{\color{black}In the fourth time slot of duration $\frac{1-\beta}{3} T$,} $\rm{R}$ will combine the decoded signals $\widetilde{s}_{{\rm{A}}}$ and $\widetilde{s}_{{\rm{B}}}$ with a \emph{power allocation ratio} $\theta\in(0,1)$ as $s_{\rm{R}}=\frac{\theta\widetilde{s}_{{\rm{A}}}+(1-\theta)\widetilde{s}_{{\rm{B}}}}{\sqrt{\theta^{2}+(1-\theta)^{2}}}$ and broadcast $s_{\rm{R}}$ to both ${\rm{A}}$ and ${\rm{B}}$ with the harvested energy $E_{\rm{total}}$.
Note that the value of $\theta$ decides how relay \rm{R} combines the
signals received from \rm{A} and \rm{B}.
{\color{black}Then the received signal at $i$ is ${y_{{\rm{R}} i}} = {h_i}\sqrt {{P_{\rm{R}}}d_i^{ - \alpha_s }} {s_{\rm{R}}} + {n_{i}}$,
where ${P_{\rm{R}}} = \frac{{{3E_{\rm{total}}}}}{{\left( {1 -\beta} \right)T}}$ is the transmit power at ${\rm{R}}$. }% and ${n_{{\rm{R}}i}}=\widetilde{n}_{{\rm{R}}} \sim {\rm{{\cal C}{\cal N}}}\left( {0,\sigma _{{\rm{R}}i}^2} \right)$ is the AWGN at node $i$.

{\color{black}For analytical simplicity, we assume $\sigma _{{\rm{R}}}^2 = \sigma _{{\rm{A}}}^2 = \sigma _{{\rm{B}}}^2 ={\sigma ^2}$ \cite{2017CL}.}
After using successive interference cancellation (SIC)\footnote{{\color{black}Since the self-signal is known at each destination node, a destination node can obtain the information from the other node by cancelling the effect of its self-signal from the received combined signal. This process requires the channel state information (CSI), which can be obtained following \cite{1603719}.}} at node $i$, the received SNR of the $\rm{R}$-$i$ link is given by
\begin{align}\label{4}
{\gamma _{{\rm{R}}i}} &= \left\{ {\begin{array}{*{20}{c}}
{\;\;\;\;\;\;\;\;\;\;\;\;\;\;\;\;\;\;0,\;\;\;\;\;\;\;\;\;\;\;\;\;\;\;\;\;\;\;\;\;{\rm{   }}{P_{\rm{in}}}{\rm{ < }}{P_{\rm{th}}}{\rm{ }}}\\
{{X_i}\left( {|{h_{\rm{A}}}{|^2}d_{\rm{A}}^{ - {\alpha _s}}{\rm{ + }}|{h_{\rm{B}}}{|^2}d_{\rm{B}}^{ - {\alpha _s}}} \right),{\rm{  }}{P_{\rm{in}}} \ge {P_{\rm{th}}}}
\end{array}} \right.
\end{align}
where {\color{black}${X_i} = \left\{ {\begin{array}{*{20}{c}}
{\frac{{3\rho \beta \eta {{\left( {1 - \theta } \right)}^2}|{h_{\rm{A}}}{|^2}d_{\rm{A}}^{ - {\alpha _s}}}}{{\left( {1 - \beta  } \right)\left[ {{\theta ^2} + {{\left( {1 - \theta } \right)}^2}} \right]}},\;{\rm{if}}\;i = {\rm{A}}}\\
{\frac{{3\rho \beta \eta {\theta ^2}|{h_{\rm{B}}}{|^2}d_{\rm{B}}^{ - {\alpha _s}}}}{{\left( {1 - \beta } \right)\left[ {{\theta ^2} + {{\left( {1 - \theta } \right)}^2}} \right]}},\;{\rm{if}}\;i = {\rm{B}}}
\end{array}} \right.$} and $\rho=\frac{P}{\sigma^2}$ is the transmit SNR.
{\color{black}By implementing the maximal ratio combining (MRC) of the signals received from nodes $\overline{i}$ and \rm{R}, the final SNR at node $i\; (i, \overline{i}\in\{\rm{A},\rm{B}\})$ is given by
\begin{align}\label{5}
%{\gamma _{i}}=\left\{ {\begin{array}{*{20}{c}}
%\;\;\;{{\gamma _{\bar i i}},\;\;\;\;{\rm{   }}{\gamma _{\bar i{\rm{R}}}} < {\gamma _{\rm{th}}}}\\
%{{\gamma _{\bar i i}} + {\gamma _{{\rm{R}} i}},{\rm{   }}{\gamma _{\bar i{\rm{R}}}} \ge {\gamma _{\rm{th}}}}
%\end{array}} \right.
{\gamma _i} = {\gamma _{{\rm{R}}i}}{\rm{ + }}{\gamma _{\bar ii}} \cdot {\mathbf{1}_{{\gamma _{\bar i{\rm{R}}}} \ge {\gamma _{{\rm{th}}}}}}
\end{align}
where ${\gamma _{\rm{th}}}$ is the predefined SNR threshold for both the relay-to-node links and the direct link and $ {\mathbf{1}_{D}}$ is the indicator function that equals 1 only when $D$ is true and 0 otherwise.}

%\vspace{-5pt}
\section{Optimal Combining and Performance Analysis}
\subsection{Optimal Combining Scheme}
Let $P_{\rm{out}}^{s}$ denote the system outage probability for the considered TS SWIPT based two-way DF relay network with TDBC.
Then $P_{\rm{out}}^{s}$ can be expressed as
\begin{align}\label{6}
&P_{\rm{out}}^{s}=1-P_{1}-P_{2}
\vspace*{-10pt}
\end{align}
where $P_{1}=\mathbb{P}\left({{\gamma _{{\rm{AB}}}} \geq \gamma _{\rm{th}}}\right)$ is the probability that the direct link achieves the given SNR threshold, $P_{2}=\mathbb{P}\big( {\gamma _{{\rm{AB}}}} < {\gamma _{{\rm{th}}}},
\min \left( {{\gamma _{{\rm{AR}}}},{\gamma _{{\rm{BR}}}}} \right) \ge {\gamma _{{\rm{th}}}},\min\left( {{\gamma _{\rm{A}}},{\gamma _{\rm{B}}}} \right) \ge {\gamma _{{\rm{th}}}},$ ${P_{\rm{in}}} \ge {P_{\rm{th}}}\big)$ is the probability of successful two-way relaying when the direct link is not available, and $\mathbb{P} \left(  \cdot  \right)$ denotes the probability.
Based on \eqref{6}, we propose an optimal combining scheme to minimize the system outage probability by optimizing the power allocation ratio $\theta$.
It can be seen that there are only two SNRs, ${\gamma _{{\rm{RA}}}}$ and ${\gamma _{{\rm{RB}}}}$, related with $\theta$.
Thus, minimizing $P_{\rm{out}}^{s}$ is equivalent to maximizing the lower SNR between ${\gamma _{{\rm{RA}}}}$ and ${\gamma _{{\rm{RB}}}}$,
and the optimization problem is formulated as
\begin{align}\label{O2}
\begin{array}{*{20}{l}}
{\mathop {{\rm{maximize}}}\limits_{\theta} \;\;\min\left({\gamma _{{\rm{RA}}}},{\gamma _{{\rm{RB}}}}\right)}\\
{{\rm{s.t.}}\;:\;0 < \theta < 1}.\!\!\!
\end{array}
\end{align}
{\color{black}By listing cases of ${\gamma _{{\rm{RA}}}}\geq{\gamma _{{\rm{RB}}}}$ and ${\gamma _{{\rm{RA}}}}\leq{\gamma _{{\rm{RB}}}}$,
it is easy to show that the optimal power allocation ratio $\theta^*$ can be obtained by letting ${\gamma _{{\rm{RA}}}}={\gamma _{{\rm{RB}}}}$. Thus, the optimal power allocation ratio $\theta^*$ is given by
\begin{align}\label{s12}
\theta^*=\frac{{|{h_{\rm{A}}}|d_{\rm{A}}^{ - \alpha_s /2}}}{{|{h_{\rm{A}}}|d_{\rm{A}}^{ - \alpha_s /2} + |{h_{\rm{B}}}|d_{\rm{B}}^{ - \alpha_s /2}}}.
\end{align}
}

{\color{black}Note that the optimal combining scheme at the relay can be extended to cases with multiple-antenna terminals. For example, for the case with multiple-antenna source terminals and a single-antenna EH relay, similar to the derivation of \eqref{s12}, the optimal power allocation ratio can be obtained as \\
$\frac{{\sqrt {{\rm{Tr}}\left( {{{\rm{H}}_{{\rm{RA}}}}{{\rm{F}}_{\rm{R}}}{\rm{F}}_{\rm{R}}^H{\rm{H}}_{{\rm{RA}}}^H} \right)} }}{{\sqrt {{\rm{Tr}}\left( {{{\rm{H}}_{{\rm{RA}}}}{{\rm{F}}_{\rm{R}}}{\rm{F}}_{\rm{R}}^H{\rm{H}}_{{\rm{RA}}}^H} \right)}  + \sqrt {{\rm{Tr}}\left( {{{\rm{H}}_{{\rm{RB}}}}{{\rm{F}}_{\rm{R}}}{\rm{F}}_{\rm{R}}^H{\rm{H}}_{{\rm{RB}}}^H} \right)} }}$, where ${\rm{F}}_{\rm{R}}$ is the beamformer at the relay and ${\rm{H}}_{{\rm{R}}i}$ denotes the channel matric from $\rm{R}$ to $i$.}
%based on the relation of ${\gamma _{RA}}$ and ${\gamma _{RB}}$, \eqref{O2} can be transformed as two cases: \textbf{Case 1:} ${\gamma _{RA}}>{\gamma _{RB}}$ and \textbf{Case 2:} ${\gamma _{RA}}\leq{\gamma _{RB}}$.

%For the case with ${\gamma _{RA}}>{\gamma _{RB}}$, we have $0<\theta<\frac{{|{h_A}|d_A^{ - \alpha_s /2}}}{{|{h_A}|d_A^{ - \alpha_s /2} + |{h_B}|d_B^{ - \alpha_s /2}}}$. Since ${\gamma _{RB}}$ monotonically increases with increasing $\theta$, the optimal power allocation ratio $\theta^*$ can be obtained by $\frac{{|{h_A}|d_A^{ - \alpha_s /2}}}{{|{h_A}|d_A^{ - \alpha_s /2} + |{h_B}|d_B^{ - \alpha_s /2}}}$.

%For the case with ${\gamma _{RA}}\leq{\gamma _{RB}}$, the range of $\theta$ is given by $\frac{{|{h_A}|d_A^{ - \alpha_s /2}}}{{|{h_A}|d_A^{ - \alpha_s /2} + |{h_B}|d_B^{ - \alpha_s /2}}}\leq\theta<1$. In this case, $\theta^*$ is also determined by $\frac{{|{h_A}|d_A^{ - \alpha_s /2}}}{{|{h_A}|d_A^{ - \alpha_s /2} + |{h_B}|d_B^{ - \alpha_s /2}}}$ due to the fact that ${\gamma _{RA}}$ decreases with the increasing of $\theta$.

%Thus, the optimal power allocation ratio $\theta^*$ is given by
%\begin{align}\label{s12}
%\theta^*=\frac{{|{h_A}|d_A^{ - \alpha_s /2}}}{{|{h_A}|d_A^{ - \alpha_s /2} + |{h_B}|d_B^{ - \alpha_s /2}}}.
%\end{align}

%\vspace*{-10pt}
\subsection{Performance Analysis}
\subsubsection{Derivation of $P_{\rm{out}}^{s}$}
In the following, we calculate $P_1$ and $P_2$ and obtain the expression of $P_{\rm{out}}^{s}$.

Firstly, based on the expression of ${\gamma _{{\rm{AB}}}}$ in \eqref{3}, $P_1$ can be computed as
\begin{align}\label{7}
P_1=\mathbb{P}\left({{\gamma _{{\rm{AB}}}} \geq \gamma _{\rm{th}}}\right)\overset{\text{(a)}}{=}\exp(-\frac{\gamma_{\rm{th}}a_t}{\rho})
\end{align}
where step (a) follows by ${\left| g \right|^2}\sim \exp(\frac{1}{\lambda_t})$ and $a_t=\frac{d_t^{\alpha_t}}{\lambda_t}$.
%\subsubsection{Derivation of $P_2$}
Then, by substituting $\theta^*$ into ${\gamma _{{\rm{RA}}}}$ and ${\gamma _{{\rm{RB}}}}$ in \eqref{4}, $P_2$ is obtained as
\begin{align}\notag\label{8}
&P_2=\mathbb{P}\bigg(Y_{\rm{A}} \ge \frac{{{\gamma _{\rm{th}}}}}{\rho },Y_{\rm{B}} \ge \frac{{{\gamma _{\rm{th}}}}}{\rho },\\
&Z < \frac{{{\gamma _{\rm{th}}}}}{\rho },Y_{\rm{A}} + Y_{\rm{B}} \ge \frac{{{P_{\rm{th}}}}}{P},Z + \Omega Y_{\rm{A}}Y_{\rm{B}} \ge \frac{{{\gamma _{\rm{th}}}}}{\rho }\bigg)
\end{align}
where $Y_i=|{h_i}{|^2}d_i^{ - \alpha_s }\;\left(i\in\{\rm{A}, \rm{B}\}\right)$, $Z=|g{|^2}d_t^{ - \alpha_t }$, and $\Omega=\frac{{3\beta \eta }}{{1 - \beta  }}$.
Since the case with $Y_{\rm{A}}\geq \frac{{{\gamma _{\rm{th}}}}}{\rho }$ and $Y_{\rm{B}}\geq \frac{{{\gamma _{\rm{th}}}}}{\rho }$ is equivalent to the case with $(Y_{\rm{A}}-\frac{{{\gamma _{\rm{th}}}}}{\rho })(Y_{\rm{B}}-\frac{{{\gamma _{\rm{th}}}}}{\rho })\geq 0$ and $Y_{\rm{A}}+Y_{\rm{B}}\geq 2\frac{{{\gamma _{\rm{th}}}}}{\rho }$, $P_2$ can be rewritten as
\begin{align}\notag\label{9}
&P_2=\mathbb{P}\left(X \ge {t_{\min }},\frac{{{\gamma _{\rm{th}}}}}{\rho } - \frac{\Omega}{Y} \le Z < \frac{{{\gamma _{\rm{th}}}}}{\rho },X \le \frac{{\rho }}{{{\gamma _{\rm{th}}}}Y} + \frac{{{\gamma _{\rm{th}}}}}{\rho }\right)=\\
&\mathbb{P}\!\left(\!{t_{\min }}\! \le\! X \le \!\frac{{\rho }}{Y{{\gamma _{\rm{th}}}}}\! +\! \frac{{{\gamma _{\rm{th}}}}}{\rho },\frac{{{\gamma _{\rm{th}}}}}{\rho } \!- \!\frac{\Omega}{Y} \le Z \!<\! \frac{{{\gamma _{\rm{th}}}}}{\rho },Y \!\le\! {\Delta _{\max }}\!\right)\!\!\!\!
\end{align}
where $X=Y_{\rm{A}}+Y_{\rm{B}}$, $Y=\frac{1}{Y_{\rm{A}}Y_{\rm{B}}}$, ${t_{\min }} = \max \left( {\frac{{{P_{\rm{th}}}}}{P},2\frac{{{\gamma _{\rm{th}}}}}{\rho }} \right)$ and ${\Delta _{\max }} =\frac{1}{\left( {{t_{\min }} - \frac{{{\gamma _{\rm{th}}}}}{\rho }} \right)\frac{{{\gamma _{\rm{th}}}}}{\rho }}$.

Let $F_Z(\Delta)$, $F_Y(\Delta)$ and $F_X(\Delta)$ denote the  cumulative distribution functions (CDFs) of $Z,Y$ and $X$, respectively. Then the expressions of $F_Z(\Delta)$, $F_Y(\Delta)$ and $F_X(\Delta)$ are obtained in \textbf{Lemma 1}.

\textbf{Lemma 1} The CDFs of $Z,Y$ and $X$ are given by
\begin{align}\notag\label{s15}
&F_{X}(\Delta)=\\
&\left\{ \begin{array}{l}
\!\!\!\!1 - {e^{ - {a_{\rm{B}}}\Delta}} - \frac{{{a_{\rm{B}}}}}{{{a_{\rm{A}}} - {a_{\rm{B}}}}}\left( {{e^{{- {a_{\rm{B}}}} \Delta}} - {e^{ { - {a_{\rm{A}}}} \Delta}}} \right),{\rm{ if}}\;{a_{\rm{A}}} \ne {a_{\rm{B}}}\\
\!\!\!\!1 - {e^{ - {a_{\rm{B}}}\Delta}} - {a_{\rm{B}}}\Delta{e^{ { - {a_{\rm{B}}}} \Delta}},{\rm{ if }}\;{a_{\rm{A}}} = {a_{\rm{B}}}
\end{array} \right.    \\
&F_{Y}(\Delta)=\frac{1}{{{\lambda _{\rm{B}}}}}\sqrt {\frac{{4{\lambda _{\rm{B}}}}d_{\rm{A}}^{\alpha_s}d_{\rm{B}}^{\alpha_s}}{{{\lambda _{\rm{A}}}\Delta}}} {K_1}\left( {\sqrt {\frac{4d_{\rm{A}}^{\alpha_s}d_{\rm{B}}^{\alpha_s}}{{{\lambda _{\rm{A}}}{\lambda _{\rm{B}}}\Delta}}} } \right)\\
&F_{Z}(\Delta)=1-\exp\left(-a_t\Delta\right)
\end{align}
where ${a_i} = \frac{{d_i^{\alpha_s} }}{{{\lambda _i}}}$ and ${K_1}\left(  \cdot  \right)$ is the modified Bessel function of the second kind.

\emph{Proof:} See the Appendix. \hfill {$\blacksquare $}

Based on Lemma 1, the probability density functions (PDFs) of $X,Y$ and $Z$, i.e., $f_X(X), f_Y(Y)$ and $f_Z(Z)$, is obtained.

Since $Z=|g{|^2}d_t^{ - \alpha_t }\geq0$, we have $P_2=P_{21}+P_{22}$, where $P_{21}$ is the probability of $\frac{{{\gamma _{\rm{th}}}}}{\rho } - \frac{\Omega}{Y}>0$ and $P_{22}$ is the probability of $\frac{{{\gamma _{\rm{th}}}}}{\rho } - \frac{\Omega}{Y}\leq0$.
%Let $F_Z(\Delta)$, $F_Y(\Delta)$ and $F_X(\Delta)$ denote the  cumulative distribution functions of $Z,Y$ and $X$, respectively. The following \textbf{Lemma. 1} can be provided to determine $F_Z(\Delta)$, $F_Y(\Delta)$ and $F_X(\Delta)$.
%$f_X(X)=\frac{{\partial {F_X}\left( X \right)}}{{\partial X}}$, $f_Y(Y)=\frac{{\partial {F_Y}\left( Y \right)}}{{\partial Y}}$ and $f_Z(Z)=\frac{{\partial {F_Z}\left( Z \right)}}{{\partial Z}}$, respectively.
For the case with $\frac{{{\gamma _{\rm{th}}}}}{\rho } - \frac{\Omega}{Y}>0$, we have $Y>\frac{\rho \Omega}{{{\gamma _{\rm{th}}}}}$. Thus, $P_{21}$ is given by
\begin{align}\notag\label{10}
&P_{21}\!\!=\!\!%\\ \notag
\mathbb{P}\!\left(\!{t_{\min }}\! \le\!\! X\!\! \le \!\frac{{\rho }}{{{Y\gamma _{\rm{th}}}}}\! \!+\! \!\frac{{{\gamma _{\rm{th}}}}}{\rho }\!,\frac{{{\gamma _{\rm{th}}}}}{\rho } \!- \!\!\frac{\Omega}{Y} \!\le \!Z \!<\! \frac{{{\gamma _{\rm{th}}}}}{\rho }\!,{\Delta _{\min }}\!\leq \!Y \!\leq\! {\Delta _{\max }}\!\!\right)\\ \notag
&=\int_{{\Delta _{\min }}}^{{\Delta _{\max }}} {{f_Y}\left( Y \right)} \int_{{t_{\min }}}^{\frac{{\rho }}{{{Y\gamma _{\rm{th}}}}} + \frac{{{\gamma _{\rm{th}}}}}{\rho }} {{f_X}\left( X\right)} \int_{\frac{{{\gamma _{\rm{th}}}}}{\rho } - \frac{\Omega}{Y}}^{\frac{{{\gamma _{\rm{th}}}}}{\rho }} {{f_Z}\left( Z \right)} dZdXdY\\
%&=\exp \left( { - \frac{{{a_t}{\gamma _{\rm{th}}}}}{\rho }} \right)\int_{{\Delta _{\min }}}^{{\Delta _{\max }}} {{f_Y}\left( Y \right)} \chi_1 \left( Y \right)dY\\
&=P_1\left[ \! {{F_Y}\left( Y \right)\!\chi_1 \left( Y \right)|_{{\Delta _{\min }}}^{{\Delta _{\max }}} \!\!- \!\!\!\underbrace{\int_{{\Delta _{\min }}}^{{\Delta _{\max }}} {{F_Y}\left( Y \right)\!\chi'_1\!\left( Y \!\right)dY}}_\Xi } \right],\!\!\!\!\!\!\!
\end{align}
where ${\Delta _{\min }}=\min\left(\frac{\rho \Omega}{{{\gamma _{\rm{th}}}}},{\Delta _{\max }}\right)$, $\chi_1 \left( Y \right) = \left[ {\exp \left( {\frac{{{a_t}\Omega }}{Y}} \right) - 1} \right]\left[ {{F_X}\left( {\frac{\rho }{{Y{\gamma _{\rm{th}}}}} + \frac{{{\gamma _{\rm{th}}}}}{\rho }} \right) - {F_X}\left( {{t_{\min }}} \right)} \right]$, and $\chi'_1\left( Y \right)=\frac{{\partial {\chi _1}\left( Y \right)}}{{\partial Y}}$.

By using Gaussian-Chebyshev quadrature approximation \cite{shi}, $\Xi$ can be approximated as follows,
\begin{align}\label{11}
\Xi\!\approx \!\frac{{\pi ({\Delta _{\max }}\! -\! {\Delta _{\min }})}}{{2M}}\sum\limits_{m = 1}^M \!{\sqrt {1 \!-\! \nu _m^2} } {F_Y}\left( {\kappa _m^{\left( 1 \right)}} \right)\chi'_1\left( {\kappa _m^{\left( 1 \right)}} \right)\!\!,
\end{align}
where $M$ is a parameter that determines the tradeoff between complexity and accuracy, ${\nu _m} = \cos \frac{{2m - 1}}{{2M}}\pi $, and $\kappa _m^{(1)} = \frac{{({\Delta _{\max }}\! -\! {\Delta _{\min }})}}{2}{\nu _m} + \frac{{({\Delta _{\max }}\! +\! {\Delta _{\min }})}}{2}$.

For the case with $\frac{{{\gamma _{\rm{th}}}}}{\rho } - \frac{\Omega}{Y}\leq0$, we have $Y\leq\frac{\rho \Omega}{{{\gamma _{\rm{th}}}}}$. Thus, $P_{22}$ is given by
\begin{align}\notag\label{12}
P_{22}&=\mathbb{P}\!\left(\!{t_{\min }}\! \le\!\! X\!\! \le \!\frac{{\rho }}{{{Y\gamma _{\rm{th}}}}}\! \!+\! \!\frac{{{\gamma _{\rm{th}}}}}{\rho }\!, 0 \!\le \!Z \!<\! \frac{{{\gamma _{\rm{th}}}}}{\rho }\!,0\!\leq \!Y \!\leq\! {\Delta _{\min }}\!\!\right)\\ \notag
&\approx\left( {1 - P_1} \right)\bigg[ {F_Y}\left( {{\Delta _{\min }}} \right){\chi _2}\left( {{\Delta _{\min }}} \right) \\
&- \frac{{\pi {\Delta _{\min }}}}{{2M}}\sum\limits_{m = 1}^M {\sqrt {1 - \nu _m^2} } {F_Y}\left( {\kappa _m^{\left( 2 \right)}} \right){{\chi'}_2}\left( {\kappa _m^{\left( 2 \right)}} \right)\bigg]
\end{align}
where $\chi_2 \left( Y \right) = {{F_X}\left( {\frac{\rho }{{Y{\gamma _{\rm{th}}}}} + \frac{{{\gamma _{\rm{th}}}}}{\rho }} \right) - {F_X}\left( {{t_{\min }}} \right)}$, $\chi'_2\left( Y \right)=\frac{{\partial {\chi _2}\left( Y \right)}}{{\partial Y}}$,
%$\left\{ {\begin{array}{*{20}{c}}{ - a_A^2\left( {\frac{\rho }{{Y{\gamma _{\rm{th}}}}} + \frac{{{\gamma _{\rm{th}}}}}{\rho }} \right)\exp \left( { - \frac{{{a_A}\rho }}{{Y{\gamma _{\rm{th}}}}} - \frac{{{a_A}{\gamma _{\rm{th}}}}}{\rho }} \right)\frac{\rho }{{{\gamma _{\rm{th}}}{Y^2}}}}\\ {\frac{{ - {a_A}{a_B}}}{{{a_A} - {a_B}}}\left[ {\exp \left( { - \frac{{{a_B}\rho }}{{Y{\gamma _{\rm{th}}}}} - \frac{{{a_B}{\gamma _{\rm{th}}}}}{\rho }} \right) - \exp \left( { - \frac{{{a_A}\rho }}{{Y{\gamma _{\rm{th}}}}} - \frac{{{a_A}{\gamma _{\rm{th}}}}}{\rho }} \right)} \right]\frac{\rho }{{{\gamma _{\rm{th}}}{Y^2}}}} \end{array}} \right.$}}
and $\kappa _m^{(2)} = \frac{{{\Delta _{\min }}\!}}{2}{\nu _m} + \frac{{\! {\Delta _{\min }}}}{2}$.

Thus, the system outage probability with the optimal combining strategy is approximated by
\begin{align}\notag\label{13}
&P_{\rm{out}}^{s}\!\approx1\! - \!{P_1}\! -\! \Theta \! +\! \frac{\pi }{{2M}}\sum\limits_{m = 1}^M \!\!\sqrt {1 - \nu _m^2} \!\bigg[ P_1\left( {{\Delta _{\max }} \!-\! {\Delta _{\min }}} \right){F_Y}\left( {\kappa _m^{\left( 1 \right)}} \right)\\
&\times{{\chi'}_1}\left( {\kappa _m^{\left( 1 \right)}} \right)
\!+ \!\left( {1 - P_1} \right){\Delta _{\min }}{F_Y}\left( {\kappa _m^{\left( 2 \right)}} \right){{\chi'}_2}\left( {\kappa _m^{\left( 2 \right)}} \right) \bigg]
\end{align}
where $
\Theta  = {P_1}\left[ {{F_Y}\left( {{\Delta _{\max }}} \right){\chi _1}\left( {{\Delta _{\max }}} \right) - {F_Y}\left( {{\Delta _{\min }}} \right){\chi _1}\left( {{\Delta _{\min }}} \right)} \right]$ $ + \left( {1 - {P_1}} \right){F_Y}\left( {{\Delta _{\min }}} \right){\chi _2}\left( {{\Delta _{\min }}} \right)$.

\subsubsection{High SNR Approximation of $P_{\rm{out}}^{s}$}
Let $\widehat{P}_{\rm{out}}^{s}$ be the high SNR approximation of $P_{\rm{out}}^{s}$.
When $\rho\rightarrow\infty$, we have $\mathop {\lim }\limits_{\rho  \to \infty } \frac{1}{\rho }{\rm{ = }}0$. In this case, we have $P_{21}\simeq 0$ due to $\Delta_{\min}\simeq \Delta_{\max}$. Further, by letting $\mathop {\lim }\limits_{\rho  \to \infty } {\chi _2}(Y) \approx {\chi _2}(Y){{\rm{|}}_{{1 \mathord{\left/
 {\vphantom {1 \rho }} \right.
 \kern-\nulldelimiterspace} \rho }{\rm{ = }}0}}{\rm{ = }}1 - {F_X}\left( {{t_{\min }}} \right)$, $\widehat{P}_{\rm{out}}^{s}$ is given by

\begin{align}\notag\label{14}
%P_{\rm{out}}^{s}&\approx
\widehat{P}_{\rm{out}}^{s}&=
1-P_1-\left( {1 - {P_1}} \right)\int_0^{{\Delta _{\min }}} {{f_Y}\left( Y \right)} \left[ {1 - {F_X}\left( {{t_{\min }}} \right)} \right]dY\\
&=\left(1-P_{1}\right)\left[1-F_Y\left(\Delta_{\min }\right)+F_Y\left(\Delta_{\min }\right)F_{X}\left({{t_{\min }}}\right)\right].
\end{align}

Although our derived expressions are complex, they provide the following advantages. Firstly, our derived expressions provide sufficiently accurate numerical evaluation of the outage performance, i.e., system outage probability and capacity.
For example, it can be observed from Fig. 1 that the results achieved by the derived expressions match the simulation result well.
Secondly, the derived expressions can be used to provide insights into the proper selection of system
parameters, such as the relay location.
%Similar insights have also been obtained in some well-cited works, such as [R1] and [R6].
Thirdly, the derived expression for the system outage probability can be used to characterize the diversity gain for SWIPT enabled two-way DF relaying networks with TDBC protocol as detailed below. {\color{blue}Note that the analysis of diversity gain has been omitted in our accepted article due to the limited space. }

The diversity gain for SWIPT enabled two-way DF
relaying networks with TDBC protocol is given by
\begin{equation}\label{1}
  d=-\mathop {\lim }\limits_{\rho  \to \infty} \frac{{ \log \left( {1 - {P_1} - {P_{21}} - {P_{22}}} \right)}}{{ \log \left( \rho  \right)}}.
\end{equation}
Since $\mathop {\lim }\limits_{\rho  \to \infty } {P_{21}} = 0$, the diversity gain can be rewritten as
\begin{align}\label{2}\notag
d&= - \mathop {\lim }\limits_{\rho  \to \infty } \frac{{ \log \left[ {\left( {1 - {P_1}} \right)\left( {1 - {F_Y}\left( {{\Delta _{\min }}} \right)\left( {1 - {F_X}\left( {{t_{\min }}} \right)} \right)} \right)} \right]}}{{ \log \left( \rho  \right)}}\\ \notag
&= - \mathop {\lim }\limits_{\rho  \to \infty } \frac{{ \log \left( {1 - {P_1}} \right)}}{{ \log \left( \rho  \right)}} - \mathop {\lim }\limits_{\rho  \to \infty } \frac{{ \log \left( {1 - {F_Y}\left( {{\Delta _{\min }}} \right)\left( {1 - {F_X}\left( {{t_{\min }}} \right)} \right)} \right)}}{{ \log \left( \rho  \right)}}\\
&\overset{\text{(a)}}{=}\underbrace{ - \mathop {\lim }\limits_{\rho  \to \infty } \frac{{ \log \left( {1 - {P_1}} \right)}}{{ \log \left( \rho  \right)}}}_{d_1} \underbrace{- \mathop {\lim }\limits_{\rho  \to \infty } \frac{{ \log \left( {{F_X}\left( {{t_{\min }}} \right)} \right)}}{{ \log \left( \rho  \right)}}}_{d_2}
\end{align}
where step (a) follows by $\mathop {\lim }\limits_{\rho  \to \infty } {{{F_Y}\left( {{\Delta _{\min }}} \right)}} = 1$.
In the following, we derive $d_1$ and $d_2$ in what follows.

(i) Derivation of $d_1$

Based on the expression of $P_1$, $d_1$  can be computed as
\begin{align}\label{3}\notag
d_1&=- \mathop {\lim }\limits_{\rho  \to \infty } \frac{{ \log \left( {1 - \exp ( - \frac{{{\gamma _{{\rm{th}}}}{a_t}}}{\rho })} \right)}}{{ \log \left( \rho  \right)}}\\
&\mathop  = \limits^{x = \frac{1}{\rho }} \mathop {\lim }\limits_{x \to 0} \frac{{\exp \left( { - {\gamma _{{\rm{th}}}}{a_t}x} \right){\gamma _{{\rm{th}}}}{a_t}x}}{{1 - \exp \left( { - {\gamma _{{\rm{th}}}}{a_t}x} \right)}}=\mathop {\lim }\limits_{x \to 0} 1 - {\gamma _{{\rm{th}}}}{a_t}x=1.
\end{align}

(ii) Derivation of $d_2$

Based on the expression of ${{F_X}\left( {{t_{\min }}} \right)}$, there are two cases for $d_2$.
For the case with ${a_{\rm{A}}} \neq {a_{\rm{B}}}$, $d_2$ can be computed as
\begin{align}\label{3-1}\notag
d_2&=- \mathop {\lim }\limits_{\rho  \to \infty } \frac{{\log \left( {1 - \frac{{{a_{\rm{A}}}}}{{{a_{\rm{A}}} - {a_{\rm{B}}}}}{e^{ - \frac{{{a_{\rm{B}}}{t_\Delta }}}{\rho }}} - \frac{{{a_{\rm{B}}}}}{{{a_{\rm{A}}} - {a_{\rm{B}}}}}{e^{ - \frac{{{a_{\rm{A}}}{t_\Delta }}}{\rho }}}} \right)}}{{\log \left( \rho  \right)}}\\ \notag
&\mathop  = \limits^{x = \frac{1}{\rho }} \mathop {\lim }\limits_{x \to 0} \frac{{{a_{\rm{A}}}{a_{\rm{B}}}{t_\Delta }x}}{{{a_{\rm{A}}} - {a_{\rm{B}}}}}\frac{{{e^{ - {a_{\rm{B}}}{t_\Delta }x}} + {e^{ - {a_{\rm{A}}}{t_\Delta }x}}}}{{1 - \frac{{{a_{\rm{A}}}}}{{{a_{\rm{A}}} - {a_{\rm{B}}}}}{e^{ - {a_{\rm{B}}}{t_\Delta }x}} - \frac{{{a_{\rm{B}}}}}{{{a_{\rm{A}}} - {a_{\rm{B}}}}}{e^{ - {a_{\rm{A}}}{t_\Delta }x}}}}\\ \notag
&=\mathop {\lim }\limits_{x \to 0} \frac{{{e^{ - {a_{\rm{B}}}{t_\Delta }x}} + {e^{ - {a_{\rm{A}}}{t_\Delta }x}} - {a_{\rm{B}}}{t_\Delta }x{e^{ - {a_{\rm{B}}}{t_\Delta }x}} - {a_{\rm{A}}}{t_\Delta }x{e^{ - {a_{\rm{A}}}{t_\Delta }x}}}}{{{e^{ - {a_{\rm{B}}}{t_\Delta }x}} + {e^{ - {a_{\rm{A}}}{t_\Delta }x}}}} \\
&= \mathop {\lim }\limits_{x \to 0} 1 - \frac{{{a_{\rm{A}}} + {a_{\rm{B}}}}}{2}{t_\Delta }x=1
\end{align}
where $t_{\min }=\frac{{{t_\Delta }}}{\rho }$ and $t_\Delta =\max \left( \frac{{{P_{\rm{th}}}}}{\sigma^{2}},2{{\gamma _{\rm{th}}}} \right)$.

Similarly, for the case with ${a_{\rm{A}}} = {a_{\rm{B}}}$, $d_2$ can be computed as
\begin{align}\label{4}
d_2= - \mathop {\lim }\limits_{\rho  \to \infty } \frac{{ \log \left( {1 - {e^{ - \frac{{{a_{\rm{B}}}{t_\Delta }}}{\rho }}} - \frac{{{a_{\rm{B}}}{t_\Delta }}}{\rho }{e^{ - \frac{{{a_{\rm{B}}}{t_\Delta }}}{\rho }}}} \right)}}{{ \log \left( \rho  \right)}}= - \mathop {\lim }\limits_{\rho  \to \infty } \frac{{\log \left( {1 - {e^{ - \frac{{{a_{\rm{B}}}{t_\Delta }}}{\rho }}}} \right)}}{{\log \left( \rho  \right)}}=1.
\end{align}

In summary, the diversity gain of the SWIPT enabled two-way DF relaying network
with TDBC protocol is given by $d=d_1+d_2=2$.

%\vspace*{-13pt}
%\subsubsection{Capacity Analysis}
%%Based on the expression of $P_{\rm{out}}^{s}$, we calculate the system outage capacity $\tau$.
%Let $\tau$ denote the system outage capacity.
%Since the effective information transmission time is $\frac{1-\beta}{3}T$, $\tau$ is given by
%\begin{small}
%\begin{align}\label{15}
%\tau=\frac{1-\beta}{3}T(1-P_{\rm{out}}^{s})U
%\end{align}
%\end{small}where $U=\log_{2}(1+\gamma_{\rm{th}})$ is the transmission rate of $\rm{A}$ and $\rm{B}$.

%\vspace*{-20pt}
\section{Simulations}
In this section, we validate the effectiveness of our proposed combining scheme and verify the accuracy of the derived expressions via $1 \times 10^{8}$ Monte-Carlo simulations.
According to \cite{7831382}, the simulation parameters are set as follows: ${d_{{\rm{A}}}} = 5\;{\rm{m}}$, ${d_{t}} = 20$m, $d_{\rm{B}}={d_{t}}-{d_{{\rm{A}}}}=15$m and $\sigma^{2}  = -70$dBm. The EH circuit sensitivity $P_{\rm{th}}$ is given by $-30$dBm and
the energy conversion efficiency is $\eta=0.6$. The transmission rate is assumed as $U=3\;{\rm{bit/s/Hz}}$ resulting in $\gamma_{\rm{th}}=2^{U}-1$. {\color{black}Unless otherwise specified, we set $\alpha_s=\alpha_t=4$ and $\beta=0.25$.} %{\color{black} Two important reference cases: (1) transmission is done only through the relay link and (2) transmission is done only through the direct link, have been included for comparison. For fair comparison, we assume that the whole energy consumption for each comparing scheme is same. Under this assumption, the transmit power for the case that transmission is done only through the direct link is computed as $\frac{2+\alpha}{3}P$.}

{\color{black}Fig. 1 plots the system outage probability versus the transmit power, where three cases are considered: (1) relay with direct link, (2) relay without direct link, and (3) direct link.
%For fair comparison, we assume that the whole energy consumption for each comparing case is same. Under this assumption, the transmit power for the case of direct link is computed as $\frac{2+\alpha}{3}P$.
Note that the case of relay with/without direct link, the power allocation ratio at relay is determined by \eqref{s12}.
For the case of relay with direct link, we use the Gaussian-Chebyshev quadrature approximation with $M=4$ and the high SNR approximation to obtain the system outage probability.
It can be observed that the result achieved by Gaussian-Chebyshev quadrature approximation matches the simulation result well, which verifies the correctness of the derived analytical expression in \eqref{13}. For the high SNR approximation, with the increasing of the transmit power, the difference between the result based on the high SNR approximation and the simulation result becomes smaller. Thus, the derived approximation in \eqref{14} is also accurate for the high SNR regions. It can also be seen that the case of relay with direct link can achieve a lower system outage probability than the cases of relay without direct link and direct link. This is because that using a relay to help the information transmission can achieve a higher diversity gain.
%match perfectly with the simulation result even $M=4$, which verifies the correctness of the derived analytical expression in \eqref{13}. For the high SNR approximation, with the increasing of the transmit power, the difference between the result based on the high SNR approximation and the simulation result becomes smaller. Thus, the derived approximation in \eqref{14} is also accurate for the high SNR regions.

Fig. 2 plots the system outage probability versus $d_{\rm{A}}$ under above three cases. For the case of relay with direct link, two schemes are considered, which are the proposed scheme and baseline scheme in which the power allocation ratio $\theta$ is fixed as $0.3, 0.5$ and $0.7$, respectively. We set $P=10$ dBm and $d_{\rm{B}}=20-d_{\rm{A}}$.}
It can be observed that with the increase of $d_{\rm{A}}$, the system outage probability increases, reaches the maximum value and then decreases. This is because that the total harvested energy is higher when the relay is closer to either of the nodes. By comparison, we can see that the proposed scheme outperforms the baseline scheme in terms of outage performance. In addition, it can also be seen that the optimal relay location is closer to either of the terminal nodes.

\begin{figure}[!t]
  \centering
  \includegraphics[width=0.45\textwidth]{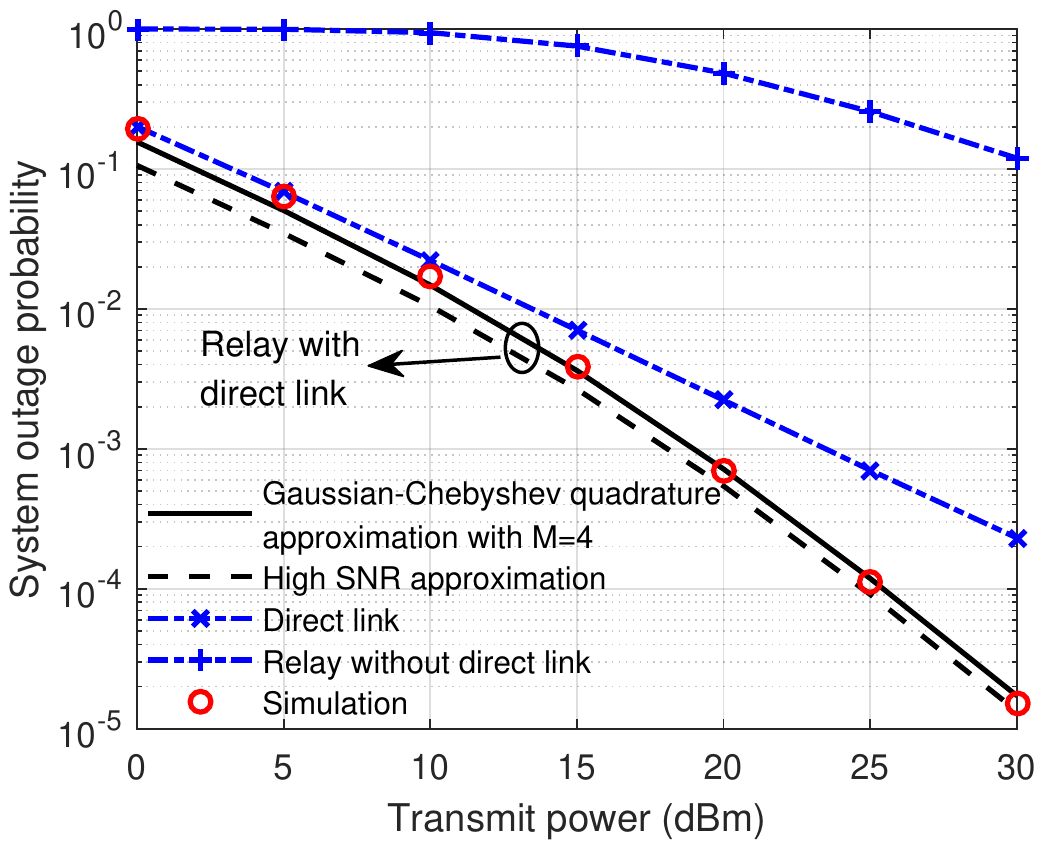}\\
  \caption{{\color{black}System outage probability versus $P$ with $\lambda_{\rm{A}}=\lambda_{\rm{B}}=1$ and $\lambda_t=2$.}}
 % \vspace*{-5pt}
\end{figure}

\begin{figure}[!t]
  \centering
  \includegraphics[width=0.45\textwidth]{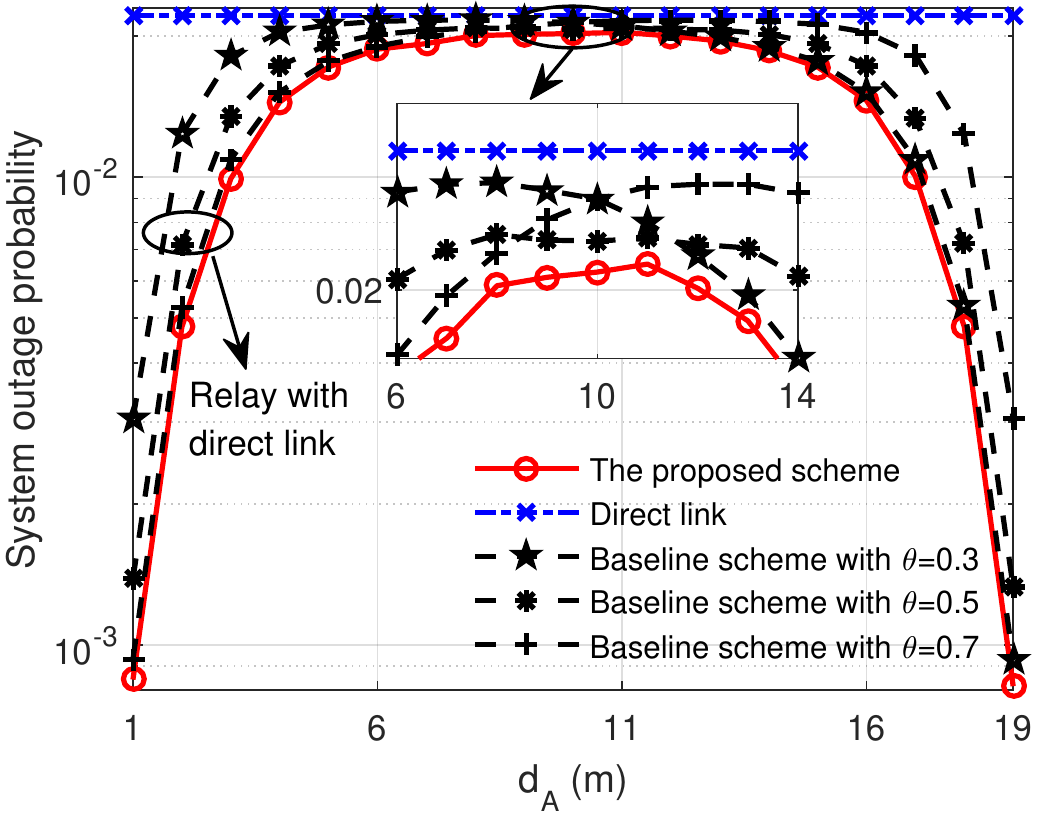}\\
  \caption{{\color{black}System outage probability versus the A-R link distance $d_{\rm{A}}$ with $\lambda_{\rm{A}}=\lambda_{\rm{B}}=1$, $\lambda_t=2$ and $P=10$ dBm.}}
 % \vspace*{-15pt}
\end{figure}
%Fig. 3 shows the system outage capacity versus TS factor for the proposed optimal combing scheme with different $\alpha_t$ configurations. {\color{black}We set $P=10$ mW, $\lambda_{\rm{A}}=\lambda_{\rm{B}}=1$ and $\lambda_t=2$. It can be seen that the capacity increases with the decreasing of $\alpha_t$, and the value of the optimal $\beta$ becomes smaller.} This is due to the fact that the influence of the direct link becomes the dominant factor to the capacity when $\alpha_t$ is small.
%{\color{black}This plot brings out the importance of selection of optimal $\beta$ for maximizing $\tau$ and shows an optimization of $\beta$ yields higher capacity.}
\section{Conclusions}
%\vspace*{-25pt}
In this paper, we have proposed an optimal combing scheme to minimize the overall system outage probability and have derived the closed-form expression for the optimal power allocation ratio $\theta$. For the proposed optimal combining scheme, we have  obtained expression of the system outage probability considering EH circuit sensitivity. We have demonstrated that both the relay location and combining scheme are critical to achieving good outage performance and our proposed combining scheme outperforms the existing combining scheme.
%We have also shown that %the optimal time for energy harvesting increases with the path loss exponent of the direct link
%While the combining scheme is important, we showed that
%an optimization of time allocation can improve the system outage capacity.

%\vspace*{-8pt}
\section*{Appendix}
%\vspace*{-15pt}
According to the definition of $F_{X}(\Delta)$, we have
\begin{align}\notag
&F_{X}(\Delta)=\mathbb{P}\left(X\leq\Delta\right)=\mathbb{P}\left[ {x \le \left( {\Delta - yd_{\rm{B}}^{ - \alpha_s }} \right)d_{\rm{A}}^{\alpha_s} ,y \le \Delta d_{\rm{B}}^{\alpha_s} } \right]\\ \notag
&=\int_0^{\Delta d_{\rm{B}}^{\alpha_s} } {\left[ {1 - \exp \left( { - {a_{\rm{A}}}\left( {\Delta - yd_{\rm{B}}^{ - {\alpha_s} }} \right)} \right)} \right]} \frac{{\exp \left( { - {y \mathord{\left/
 {\vphantom {y {{\lambda _{\rm{B}}}}}} \right.
 \kern-\nulldelimiterspace} {{\lambda _{\rm{B}}}}}} \right)}}{{{\lambda _{\rm{B}}}}}dy\\
 &={1 - {e^{ - {a_{\rm{B}}}\Delta}} - \frac{{{e^{ - {a_{\rm{B}}}\Delta}}}}{{{\lambda _{\rm{B}}}}}\int_0^{\Delta d_{\rm{B}}^{\alpha_s} } {\exp \left( {\frac{{{a_{\rm{A}}} - {a_{\rm{B}}}}}{{{a_{\rm{B}}}{\lambda _{\rm{B}}}}}y} \right)} dy}\!\!
\end{align}
where $x=|{h_{\rm{A}}}{|^2}$, $y=|{h_{\rm{B}}}{|^2}$ and ${a_i} = \frac{{d_i^{\alpha_s} }}{{{\lambda _i}}}$.
When ${a_{\rm{A}}}={a_{\rm{B}}}$, (25) can be computed as $F_{X}(\Delta)={1 - {e^{ - {a_{\rm{B}}}\Delta}} - {a_{\rm{B}}}\Delta{e^{ - {a_{\rm{B}}}\Delta}}}$.
%\begin{small}
%\begin{align}
%F_{t_2}(t)={1 - {e^{ - {a_B}t}} - {a_B}t{e^{ - {a_A}t}}}.
%\end{align}
For the case with ${a_{\rm{A}}}\neq{a_{\rm{B}}}$, (25) is given by
\begin{align}
F_{X}(\Delta)={1 - {e^{ - {a_{\rm{B}}}\Delta}} - \frac{{{a_{\rm{B}}}}}{{{a_{\rm{A}}} - {a_{\rm{B}}}}}\left( {{e^{ - {a_{\rm{B}}}\Delta}} - {e^{ - {a_{\rm{A}}}\Delta}}} \right)}.
\end{align}
Thus, $F_{X}(\Delta)$ can be rewritten as (12).

%\subsection{Derivation of $F_{t_3}(t)$}
Similarly, $F_{Y}(\Delta)$ and $F_{Z}(\Delta)$ are given by
\begin{align}\notag
&F_{Y}(\Delta)=\mathbb{P}(x\geq\frac{d_{\rm{A}}^{\alpha_s}d_{\rm{B}}^{\alpha_s}}{y\Delta})=\frac{1}{{{\lambda _{\rm{B}}}}}\int_0^{ + \infty } {\exp \left( { - \frac{d_{\rm{A}}^{\alpha_s}d_{\rm{B}}^{\alpha_s}}{{{\lambda _{\rm{A}}}\Delta y}} - \frac{y}{{{\lambda _{\rm{B}}}}}} \right)dy}\\
&=\frac{1}{{{\lambda _{\rm{B}}}}}\sqrt {\frac{{4{\lambda _{\rm{B}}}d_{\rm{A}}^{\alpha_s}d_{\rm{B}}^{\alpha_s}}}{{{\lambda _{\rm{A}}}\Delta}}} {K_1}\left( {\sqrt {\frac{4d_{\rm{A}}^{\alpha_s}d_{\rm{B}}^{\alpha_s}}{{{\lambda _{\rm{A}}}{\lambda _{\rm{B}}}\Delta}}} } \right)\\
&F_{Z}(\Delta)=\mathbb{P}(Z\leq\Delta)=\mathbb{P}(|g{|^2}\leq\Delta d_t^{  \alpha_t })=1-\exp\left(-a_t\Delta\right)
\end{align}
where ${K_1}\left(  \cdot  \right)$ is the modified Bessel function of the second kind.
The proof is completed.
%\vspace*{-5pt}
\ifCLASSOPTIONcaptionsoff
  \newpage
\fi
\bibliographystyle{IEEEtran}
\bibliography{refa}

\begin{thebibliography}{10}
\providecommand{\url}[1]{#1}
\csname url@samestyle\endcsname
\providecommand{\newblock}{\relax}
\providecommand{\bibinfo}[2]{#2}
\providecommand{\BIBentrySTDinterwordspacing}{\spaceskip=0pt\relax}
\providecommand{\BIBentryALTinterwordstretchfactor}{4}
\providecommand{\BIBentryALTinterwordspacing}{\spaceskip=\fontdimen2\font plus
\BIBentryALTinterwordstretchfactor\fontdimen3\font minus
  \fontdimen4\font\relax}
\providecommand{\BIBforeignlanguage}[2]{{%
\expandafter\ifx\csname l@#1\endcsname\relax
\typeout{** WARNING: IEEEtran.bst: No hyphenation pattern has been}%
\typeout{** loaded for the language `#1'. Using the pattern for}%
\typeout{** the default language instead.}%
\else
\language=\csname l@#1\endcsname
\fi
#2}}
\providecommand{\BIBdecl}{\relax}
\BIBdecl

\bibitem{7744827}
W.~Guo \emph{et~al.}, ``Simultaneous information and energy flow for {IoT}
  relay systems with crowd harvesting,'' \emph{IEEE Commun. Mag.}, vol.~54,
  no.~11, pp. 143--149, November 2016.

\bibitem{7831382}
S.~Modem and S.~Prakriya, ``Performance of analog network coding based two-way
  {EH} relay with beamforming,'' \emph{IEEE Trans. Commun.}, vol.~65, no.~4,
  pp. 1518--1535, April 2017.

\bibitem{7807356}
T.~P. Do \emph{et~al.}, ``Simultaneous wireless transfer of power and
  information in a decode-and-forward two-way relaying network,'' \emph{IEEE
  Trans. Wireless Commun.}, vol.~16, no.~3, pp. 1579--1592, March 2017.

\bibitem{7876801}
J.~Rostampoor, S.~M. Razavizadeh, and I.~Lee, ``Energy efficient precoding
  design for {SWIPT} in {MIMO} two-way relay networks,'' \emph{IEEE Trans. Veh.
  Technol.}, vol.~66, no.~9, pp. 7888--7896, Sept 2017.

\bibitem{7037438}
Y.~Liu, L.~Wang, M.~Elkashlan \emph{et~al.}, ``Two-way relaying networks with
  wireless power transfer: Policies design and throughput analysis,'' in
  \emph{Proc. IEEE Globecom}, Dec 2014, pp. 4030--4035.

\bibitem{8287997}
Z.~Wang \emph{et~al.}, ``Dynamic power splitting for three-step two-way
  multiplicative {AF} relay networks,'' in \emph{IEEE VTC}, Sept 2017, pp.
  1--5.

\bibitem{8361446}
Y.~Ye, Y.~Li, Z.~Wang, X.~Chu, and H.~Zhang, ``Dynamic asymmetric power
  splitting scheme for {SWIPT} based two-way multiplicative {AF} relaying,''
  \emph{IEEE Signal Process. Lett.}, pp. 1--1, 2018.

\bibitem{2017CL}
N.~T.~P. Van, S.~F. Hasan, X.~Gui, S.~Mukhopadhyay, and H.~Tran, ``Three-step
  two-way decode and forward relay with energy harvesting,'' \emph{IEEE Commun.
  Lett.}, vol.~21, no.~4, pp. 857--860, April 2017.

\bibitem{shi}
L.~Shi, W.~Cheng, Y.~Ye, H.~Zhang, and R.~Q. Hu, ``Heterogeneous
  power-splitting based two-way {DF} relaying with non-linear energy
  harvesting,'' in \emph{Proc. Globecom}, Dec 2018, pp. 1--7, to appear.

\bibitem{8377371}
D.~S. Gurjar, U.~Singh, and P.~K. Upadhyay, ``Energy harvesting in hybrid
  two-way relaying with direct link under nakagami-m fading,'' in \emph{Proc.
  IEEE WCNC}, April 2018, pp. 1--6.

\bibitem{8103768}
F.~Jameel, S.~Wyne, and Z.~Ding, ``Secure communications in three-step two-way
  energy harvesting {DF} relaying,'' \emph{IEEE Commun. Lett.}, vol.~22, no.~2,
  pp. 308--311, Feb 2018.

\bibitem{8364583}
A.~Mukherjee, T.~Acharya, and M.~R.~A. Khandaker, ``Outage analysis for
  {SWIPT}-enabled two-way cognitive cooperative communications,'' \emph{IEEE
  Trans. Veh. Technol.}, vol.~67, no.~9, pp. 9032--9036, Sept 2018.

\bibitem{5738653}
Z.~Yi, M.~Ju, and I.~M. Kim, ``Outage probability and optimum combining for
  time division broadcast protocol,'' \emph{IEEE Trans. Wireless Commun.},
  vol.~10, no.~5, pp. 1362--1367, May 2011.

\bibitem{1603719}
A.~Bletsas, A.~Khisti, D.~P. Reed, and A.~Lippman, ``A simple cooperative
  diversity method based on network path selection,'' \emph{IEEE J. Sel. Areas
  Commun.}, vol.~24, no.~3, pp. 659--672, March 2006.

\end{thebibliography}

\end{document}